\documentclass[twocolumn]{revtex4}[12pt]
\draft
\begin{document}
\title{Growing electrostatic modes in the isothermal pair plasma of the pulsar magnetosphere}
\author{U. A. Mofiz$^{1}$\footnote{Corresponding author; $\:$ email address: mofiz@bracu.ac.bd telephone:880-2-8824051-4 ext.4078, fax: 880-2-8810383.} , M. R. Amin$^2$, and P. K. Shukla$^3$}
\address{$^1$Department of Mathematics and Natural Sciences, BRAC University, 66 Mohakhali, Dhaka-1212, Bangladesh\\
$^2$Department of Electronics and Communication Engineering, East West University, 43 Mohakhali , Dhaka-1212, Bangladesh\\
$^3$Institut f\"{u}r Theoretische Physik IV, Fak\"{u}ltat f\"{u}r Physik and Astronomie, Ruhr-Universit\"{a}t Bochum, D-44780 Bochum, Germany}
\vspace{1in}
\begin{abstract}
{\bf{Abstract}}\\
It is shown that a  strongly  magnetized isothermal pair plasma near the surface of a pulsar  supports  low-frequency  (in comparison to electron cyclotron frequency) toroidal electrostatic plasma modes in the equatorial region. Physically, the thermal pressure coupled with the magnetic pressure creates the low frequency oscillations which may grow for particular case of inhomogeneities of the equilibrium magnetic field and the pair plasma density.
\end{abstract}
\maketitle
{\bf{1 Introduction}}\\ \\
Study of pair plasma in the pulsar magnetosphere is related to the investigation of radio emission coming
from these sources (Buzzi and Hines, 1995; Mofiz, 1997 and the references therein). Radio pulsars, which are rotating neutron stars with spin periods ranging from 1.57 ms to 5 ms, are characterized by surface magnetic fields in the order of $10^{12} G$, radii of about 10 km. A rotating magnetized neutron star generates huge potential difference between different parts of its surface (Goldreich and Julian, 1969). Theoretical models (Sturrock and Baker 1979; Ruderman and Sutherland, 1975; Arons and Scharlemann, 1979) have been developed to predict the production of pair plasma (electron-positron) in the pulsar magnetosphere. A mechanism of cascade generation of particles has been suggested, according to which secondary electrons and positrons result from pair production induced by high-energy curvature radiation photons- emitted by primary electron beams coming from pulsar surface. Pair plasmas are composed of  charged particles with same mass and opposite charges, which admit the time and space scales that significantly differs from those of an electron - ion plasma .  There appear a great variety of linear and nonlinear plasma modes in the strongly magnetized pair plasma. A good number of works are done by several authors ( Sakai and Kawata, 1980; Lominadze $\it{et}$  $\it {al.}$, 1983; Mofiz, 1989, 1990, 1997;Mofiz $\it{et}$  $\it {al.}$1985, 1987;   Shukla $\it{et}$  $\it {al.}$,2007). Recent discoveries (Sana $\it{et}$  $\it {al.}$,2010) of kilohertz quasi-periodic oscillations (kHz QPOs)  may also be related with the typical dynamical time scales of pair plasma close to the surface of the neutron star.\\

In this paper, we have investigated the low frequency (in comparison to the electron cyclotron frequency) electrostatic oscillations in the equatorial region of the pulsar magnetosphere. We found that thermal pressure coupled with the magnetic pressure creates the low frequency oscillations which propagate along the toroidal direction of the equatorial plasma. The magnetic field inhomogeneities and the equilibrium plasma density in particular situations may admit growing low frequency electrostatic modes in the pulsar magnetosphere. \\ \\
{\bf{ 2 Formalism: Equilibrium Magnetosphere}}\\ \\
{\bf{A. Equilibrium magnetic field}}\\
To have an idea of the equilibrium plasma density in the pulsar magnetosphere, we consider the simple model of dipolar nonalighned magnetic field ${\bf B}=\{B_{r},B_{\theta},0\}$ of the pulsar  (Ginzburg and Ozernoy, 1964; Muslimov and Harding, 1997; Mofiz and Ahmedov,2000; Prasanna et. al., 1989 ) with the components:

\begin{eqnarray}
B_r(r,\theta)=B_p\left(\frac{R}{r}\right)^3\cos{\theta},\label{beq1}
\end{eqnarray}
\begin{eqnarray}
B_\theta(r,\theta)=\frac{B_p}{2}\left(\frac{R}{r}\right)^3\sqrt{1-\frac{r_g}{r}}sin\theta,\label{beq2}
\end{eqnarray}
where, $B_p=\frac{\mu}{R^3}$ is the Newtonian value of the magnetic field at the pole of the star with the magnetic moment $\mu$  and radius $R$; $r_g$ is the Schwarzschild radius of the pulsar( Shapiro and Teukolsky,2007).{\bf The detail calculation of magnetic field structure around a nonaligned neutron star was given by Ginzburg and Ozernoy in 1964, which was further developed by different authors including Muslimov and Harding in 1997. Here, we consider the expression for the components of magnetic field $B_r,B_\theta$, which are approximately valid near the surface of a pulsar}\\

Introducing a compactness parameter $\epsilon=\frac{r_g}{R}$, dimensinless magnetic field $\bar B_0(\bar r,\theta)=\frac{B_0(r,\theta)}{B_p}$ and dimensionless radial distance $\bar r=\frac{r}{R}$, we can write the magnitude of the pulsar magnetic field as:
\begin{eqnarray}
\bar B_0(r,\theta)=\left(\frac{1}{\bar r}\right)^3\sqrt{\cos^2{\theta}+\frac{1}{4}(1-\frac{\epsilon}{\bar r})sin^2{\theta}}.\label{beq3}
\end{eqnarray}

The magnitudes of the magnetic field at the pole and at the equatorial region of the pulsar magnetosphere, are shown in Fig.[1] The figure shows that the magnetic field is more strong at the pole , where gravity has no effect and it is less at the equator where gravity reduces the field. \\

\begin{figure} \vspace{5.6cm}
\includegraphics{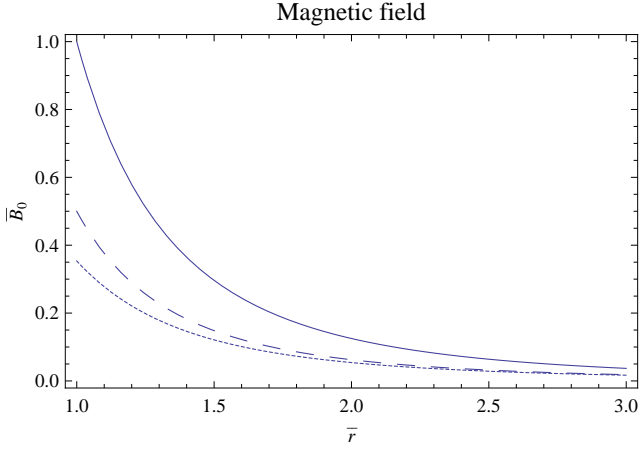} \caption{Magnetic field at the pole and at the equatorial region of the pulsar magnetosphere : solid line (---) at the pole; broken line ($-$ $-$) and doted line (...) at the equator for $\epsilon = 0, .5 $, respectively.  }
\end{figure}

To see the magnetic field configuration, we solve the equation for magnetic lines of force (Banarjee \emph{et al}., 1995)
\begin{eqnarray}
\frac{dr}{B_r}=\frac{rd\theta}{B_\theta},\label{beq4}
\end{eqnarray}
which gives
\begin{eqnarray}
2r\left(1+\sqrt{1-\frac{r_g}{r}}\right)-r_g=4C\exp^{{2(\sqrt{1-\frac{r_g}{r}}-1)}} sin^2\theta,\label{beq5}
\end{eqnarray}
where $C$ is a constant of integration. \\

Eq.(5) can be written in dimensionless form as

\begin{eqnarray}
2{\bar r}(1+\sqrt{1-\frac{\epsilon}{\bar r}})-\epsilon=4\bar C\exp^{{2(\sqrt{1-\frac{\epsilon}{\bar r}}-1)}} sin^2\theta,\label{beq6}
\end{eqnarray}

where, we put $\bar C=\frac{C}{R}$. \\ \\

For Newtonian case (no gravity) $\epsilon=0$, we have
\begin{eqnarray}
\bar r=\bar C sin^2\theta,\label{beq7}
\end{eqnarray}

Now, in order to visualize the field line structure, it is useful to transform over to a Cartesian frame through the usual relation ( $X=rsin\theta cos\Phi,Y=rsin\theta sin\Phi,Z=rcos\theta$). Then, from Eq.(7) the magnetic lines of force projected on the meridional plane $(\Phi=0)$ are described by the equation

\begin{eqnarray}
\bar z^2=(\bar C\bar x ^2)^{2/3}-\bar x^2,\label{beq8}
\end{eqnarray}
 where, $\bar x =\frac{x}{R},\bar z =\frac{z}{R}$. \\ \\

For compact object with gravity
 ($\epsilon \neq 0)$, the equation for lines of force [ Eq.(6)] can be written as
\begin{eqnarray}
\bar r^3=\frac{4\bar C\bar x^2\exp^{2(\sqrt{1-\frac{\epsilon}{\bar r}}-1)}+\epsilon \bar r^2}{2(1+\sqrt{1-\frac{\epsilon}{\bar r}})},\label{beq9}
\end{eqnarray}

which for $\frac{\epsilon}{\bar r}\ll 1$, can be written as

\begin{eqnarray}
\bar r^3=\frac{4\bar C\bar x^2\exp^{2(\sqrt{1-\frac{\epsilon}{\bar r_x}}-1)}+\epsilon \bar r_x^2}{2(1+\sqrt{1-\frac{\epsilon}{\bar r_x}})},\label{beq9}
\end{eqnarray}

where, we put $\frac{\epsilon}{\bar r}\sim \frac{\epsilon}{\bar r_x}$ with $\bar r_x=\sqrt[3]{\bar C \bar x^2}$. \\

Then, the magnetic lines of force in the meridional plane are described by the equation

\begin{eqnarray}
\bar z^2=\left\{\frac{4\bar C\bar x^2\exp^{2(\sqrt{1-\frac{\epsilon}{\bar r_x}}-1)}+\epsilon \bar r_x^2}{2(1+\sqrt{1-\frac{\epsilon}{\bar r_x}})}\right\}^{2/3}-\bar x^2,\label{beq10}
\end{eqnarray}

Using Eqs. (8) and (10), we plot the lines of forces for Newtonian case ($\epsilon =0$) and for gravitational case ($\epsilon =.5$) putting the values of $\bar C=1,2,3,5$; respectively, which are shown in Fig.[2] \\ \\
\begin{figure} \vspace{5.5cm}
\includegraphics{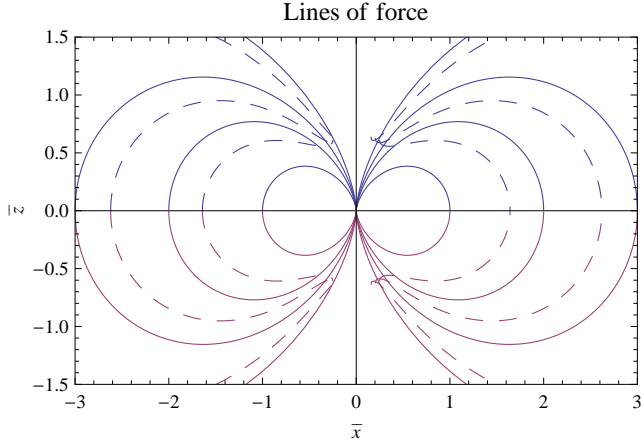} \caption{ Magnetic lines of force in the meridional plane $(\bar x,\bar z)$ of the pulsar magnetosphere:solid line (---)[Newtonian case $\epsilon =0$],broken line($-$ $-$)[gravitational case,$\epsilon =.5$]. }
\end{figure}

To see the visual observation  of the ambient magnetic field around a pulsar, we perform the density plot of $\bar B_0$ from the Eq.(3) in the meridional plane $(\bar x,\bar z)$ which is  described by the equation

\begin{eqnarray}
\bar B_0(\bar x,\bar z)=(\frac{1}{\bar x^2+\bar z^2})^{3/2}\sqrt{\frac{\bar z^2}{\bar x^2+\bar z^2}+\frac{1}{4}(1-\frac{\epsilon}{\sqrt{\bar x^2+\bar z^2}})\frac{\bar x^2}{\bar x^2+\bar z^2} } \label{beq11}
\end{eqnarray}

The distribution of magnetic field magnitudes are shown in Fig.[3] and Fig.[4] for the Newtonian and gravitational cases, respectively. \\ \\

\begin{figure} \vspace{6.5cm}
\includegraphics{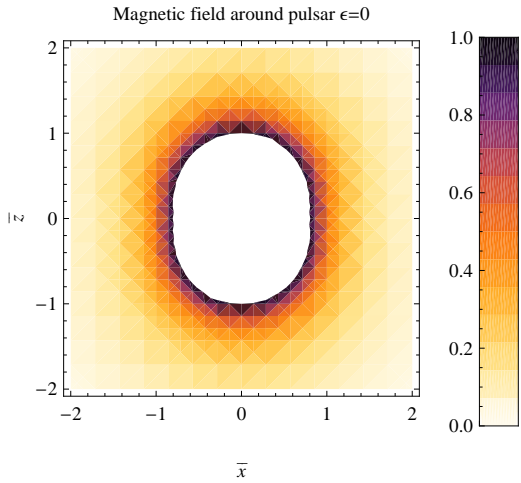} \caption{ Distribution of magnetic field magnitudes in the meridional plane $(\bar x,\bar z)$of the pulsar magnetosphere:(Newtonian case,  $\epsilon =0$). }
\end{figure}

\begin{figure} \vspace{6.5cm}
\includegraphics{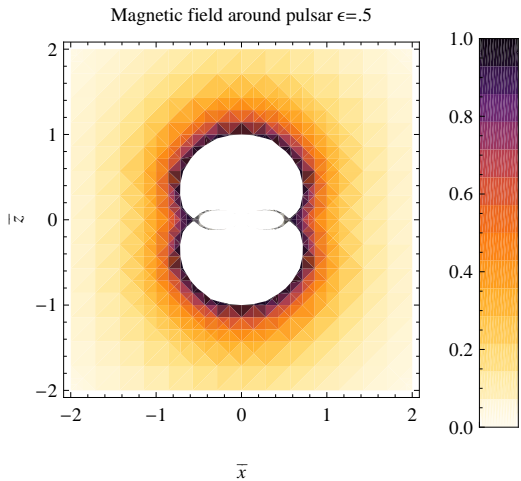} \caption{ Distribution of magnetic field magnitudes in the meridional plane $(\bar x,\bar z)$of the pulsar magnetosphere:(gravitational case$\epsilon =.5$). }
\end{figure}

{\bf{B. Equilibrium plasma density.}}\\

Now, we study the equilibrium plasma density around pulsar. The equilibrium is due to the balance of magnetic pressure and the thermal pressure, which is described  by the equation (Krall and Trivelpiece,1973)

\begin{eqnarray}
\nabla\left(\frac{{B}_0^2(r,\theta)}{8\pi}+n_0(r,\theta)T_0\right)=\frac{\bf{B}_0\cdot\nabla \bf{B}_0}{8\pi}. \label{beq11}
\end{eqnarray}
In spherical coordinate system:
\begin{eqnarray}
({\bf{B}_0}\cdot\nabla {\bf{B}_0})_r=\frac{1}{2}\frac{\partial B_r^2}{\partial r}-\frac{B_\theta^2}{r}(\frac{2}{\sqrt{1-\frac{r_g}{r}}}+1),\label{beq12}
\end{eqnarray}
\begin{eqnarray}
({\bf{B}_0}\cdot\nabla {\bf{B}_0})_\theta=\frac{1}{r}\frac{\partial B_\theta^2/2}{\partial\theta}-\frac{2B_rB_\theta}{r}(1-\frac{r_g}{4r(1-\frac{r_g}{r})}).\label{beq13}
\end{eqnarray}
Then, Eq.(13) yields:

\begin{eqnarray}
\frac{B_0^2(r,\theta)}{8\pi}+ 2n_0(r,\theta)T_0&=& -\frac{1}{4\pi}\int \frac{B_\theta^2}{r}(\frac{2}{\sqrt{1-\frac{r_g}{r}}}+1)dr \nonumber \\
&-&\frac{1}{2\pi}\int B_rB_\theta (1-\frac{r_g}{4r(1-\frac{r_g}{r})})d\theta, \label{beq16}
\end{eqnarray}
Now, we investigate the equilibrium equation (16) for the following two cases: \\ \\

{\bf{i) Newtonian case}}\\ \\

For a Newtonian case, $r_g=0$, so\\

$B_r(r,\theta)=B_p\left(\frac{R}{r}\right)^3\cos{\theta}$,
$B_\theta(r,\theta)=\frac{B_p}{2}\left(\frac{R}{r}\right)^3sin\theta$ \\

Then from Eq. (16), we have
\begin{eqnarray}
\frac{B_0^2(r,\theta)}{8\pi}+ 2n_0(r,\theta)T_0&=& -\frac{3B_p^2}{16\pi}R^6sin^2\theta\left \{-\frac{1}{6r^6}+C_1(\theta)\right\}\nonumber \\ &-&\frac{B_p^2}{4\pi}(\frac{R}{r})^6\left\{\frac{sin^2\theta}{2}+C_2(r)\right\}, \label{beq17}
\end{eqnarray}

where $C_1(\theta)$ and $C_2(r)$ are the constants of integrations, which are to be determined from the boundary conditions.\\

We set the boundary conditions:   i)at $r\rightarrow\infty;$ $B_0\rightarrow 0$ and $n_0\rightarrow 0$, ii) at $r = R$ and $\theta=0$;  $ n_0 \rightarrow n_p$, where, $n_p$ is the equilibrium plasma density at the pole of the pulsar. Then, we find
$C_1(\theta)=0$ and $C_2(r)=const=-4\pi(1+\frac{2n_pT_0}{B_P^2}).$
Thus, from Eq.(17), we get
\begin{eqnarray}
n_0(r,\theta)\equiv n_0(r)=n_p(\frac{R}{r})^6, \label{beq18}
\end{eqnarray}

which is independent of $\theta$ and can be represented as
 \begin{eqnarray}
\bar n_0(\bar r)= (\frac{1}{\bar r})^6, \label{beq19}
\end{eqnarray}
 where, $\bar n_0=\frac{n_0}{n_p}$.
 It shows that in the Newtonian case, the plasma density around the pulsar is isotropic and is a steeply decreasing function of radial distance, which is shown in  Fig.[5]. The corresponding density distribution in the meridional plane ($\bar x,\bar z $) is shown in Fig.[6]. \\ \\

\begin{figure} \vspace{6.5cm}
\includegraphics{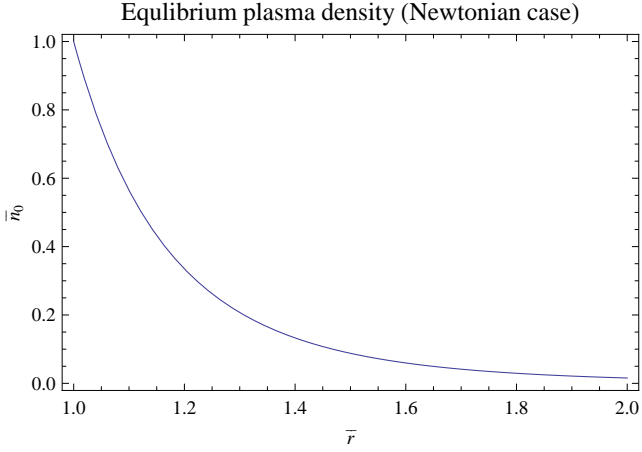} \caption{Equilibrium plasma density $\bar n_0(\bar r)$(Newtonian case, $\epsilon =0$) }
\end{figure}

\begin{figure} \vspace{6.5cm}
\includegraphics{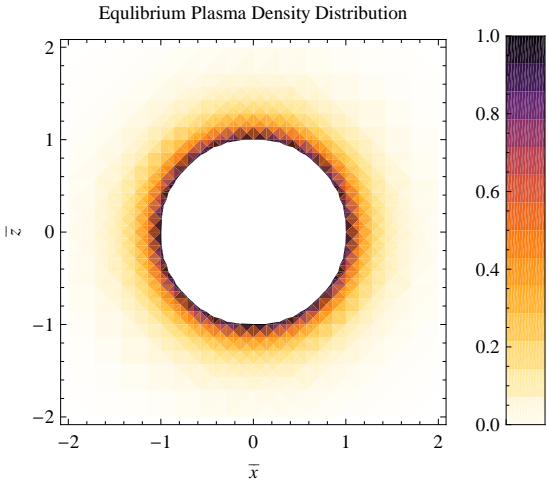} \caption{Equilibrium plasma density distribution in the meridional plane $(\bar x, \bar z)$ (Newtonian case,$\epsilon =0$ ) }
\end{figure}

{\bf{ ii) Gravitational  case}}\\ \\

In the case of gravity $r_g\neq 0$, so

$B_r(r,\theta)=B_p\left(\frac{R}{r}\right)^3\cos{\theta}$,\\

$B_\theta(r,\theta)=\frac{B_p}{2}\left(\frac{R}{r}\right)^3\sqrt{1-\frac{r_g}{r}}sin\theta$.\\

Then, from Eq.(16) we find

\begin{eqnarray}
\frac{{B}_0^2(r,\theta)}{8\pi}+ 2n_0(r,\theta)T_0&=& -\frac{B_p^2}{16\pi}(\frac{R}{r})^6sin^2\theta\{-\frac{1}{6}+\frac{r_g}{7r} \nonumber \\
&+&\frac{4}{9009}(\frac{r}{r_g})^6\sqrt{1-\frac{r_g}{r}}P(\frac{r_g}{r})\nonumber \\ &+&(\frac{r}{r_g})^6D_1(\theta)\} \nonumber \\
&-&\frac{B_p^2}{4\pi}(\frac{R}{r})^6\sqrt{1-\frac{r_g}{r}}\left(1-\frac{r_g}{4r(1-\frac{r_g}{r})}\right)\nonumber \\
&\times&\left\{\frac{sin^2\theta}{2}+D_2(r)\right\}, \label{beq20}
\end{eqnarray}

where $D_1(\theta)$ and $D_2(r)$ are the constants of integration and
\begin{eqnarray}
P(\frac{r_g}{r})&=&256+128\frac{r_g}{r}+96(\frac{r_g}{r})^2+80(\frac{r_g}{r})^3 \nonumber\\
&+&70(\frac{r_g}{r})^4 +63(\frac{r_g}{r})^5-693(\frac{r_g}{r})^6. \label{beq21}
\end{eqnarray}

The constants $D_1(\theta)$ and $D_2(r)$ may be determined from the earlier boundary conditions, i.e.:i) at $r\rightarrow \infty;$ $B_0\rightarrow 0$ and $n_0\rightarrow 0$ and ii) at $r=R$ and $ \theta=0$; $n_0=n_p$, where $n_p$ is the equilibrium plasma density at the pole. Thus, we find
\begin{eqnarray}
D_1(\theta)=const.=-\frac{4\times 256}{9009}. \label{beq22}
\end{eqnarray}
\begin{eqnarray}
D_2(r)=const.=-\frac{1}{\sqrt{1-\frac{r_g}{R}}(1-\frac{r_g}{4R(1-\frac{r_g}{R})})}
(1+\frac{2n_pT_0}{B_p^2/8\pi}),\label{beq23}
\end{eqnarray}

Then, from Eq.(20), we obtain the equilibrium plasma density under gravity as

\begin{eqnarray}
n_0(r,\theta)&=&n_p(\frac{R}{r})^6\left[\frac{\sqrt{1-\frac{r_g}{r}}(1-\frac{r_g}{4r(1-\frac{r_g}{r})})}
{\sqrt{1-\frac{r_g}{R}}
(1-\frac{r_g}{4R(1-\frac{r_g}{R})})}\right] \nonumber \\
&-&\frac{B_p^2}{16\pi T_0}[\left\{cos^2\theta + sin^2\theta\left (\frac{1}{4}(1-\frac{r_g}{r})\right )\right\}\nonumber \\
&+&sin^2\theta\sqrt{1-\frac{r_g}{r}}\left(1-\frac{r_g}{4r(1-\frac{r_g}{r})}\right) \nonumber \\
&+&\frac{1}{2}sin^2\theta\{-\frac{1}{6}+\frac{r_g}{7r}+\frac{4}{9009}(\frac{r}{r_g})^6 \nonumber \\
&\times&\left(\sqrt{1-\frac{r_g}{r}}P(\frac{r_g}{r})-256\right)\}\nonumber \\
&-&\frac{\sqrt{1-\frac{r_g}{r}}(1-\frac{r_g}
{4r(1-\frac{r_g}{r})})}
{\sqrt{1-\frac{r_g}{R}}
(1-\frac{r_g}{4R(1-\frac{r_g}{R})})}],\label{beq24}
\end{eqnarray}

\vspace{.5cm}

which can be written in dimensionless form as

\begin{eqnarray}
\bar n_0(\bar r,\theta)&=&(\frac{1}{\bar r})^6\left[\frac{\sqrt{1-\frac{\epsilon }{\bar r}}(1-\frac{\epsilon}{4\bar r(1-\frac{\epsilon}{\bar r})})}
{\sqrt{1-\frac{\epsilon}{}}
(1-\frac{\epsilon}{4(1-\epsilon)})}\right] \nonumber \\
&-&\frac{\bar B_p^2}{8\pi }[\left\{cos^2\theta + sin^2\theta\left (\frac{1}{4}(1-\frac{\epsilon }{\bar r})\right )\right\}\nonumber \\
&+&sin^2\theta\sqrt{1-\frac{\epsilon }{\bar r}}\left(1-\frac{\epsilon}{4\bar r(1-\frac{\epsilon }{\bar r})}\right) \nonumber \\
&+&\frac{1}{2}sin^2\theta\{-\frac{1}{6}+\frac{\epsilon}{7\bar r}+\frac{4}{9009}(\frac{\bar r}{\epsilon})^6 \nonumber \\
&\times&\left(\sqrt{1-\frac{\epsilon }{\bar r}}P(\frac{\epsilon }{\bar r})-256\right)\}\nonumber \\
&-&\frac{\sqrt{1-\frac{\epsilon }{\bar r}}(1-\frac{\epsilon }
{4\bar r(1-\frac{\epsilon }{\bar r})})}
{\sqrt{1-\epsilon}
(1-\frac{\epsilon}{4(1-\epsilon)})}]. \label{beq25}
\end{eqnarray}

Here, we introduce the dimensionless quantities:
$\bar n_0(\bar r,\theta)=\frac{n_0(\bar r,\theta)}{n_p}$ and $\bar B_p^2=\frac{B_p^2}{2n_pT_0}.$ \\

Now, for different values of compactness parameter $\epsilon =.1, .5, .7$,respectively; we plot the equilibrium plasma density in the equatorial disc of the pulsar magnetosphere which are shown in Fig.[7].

\begin{figure} \vspace{5.5cm}
\includegraphics{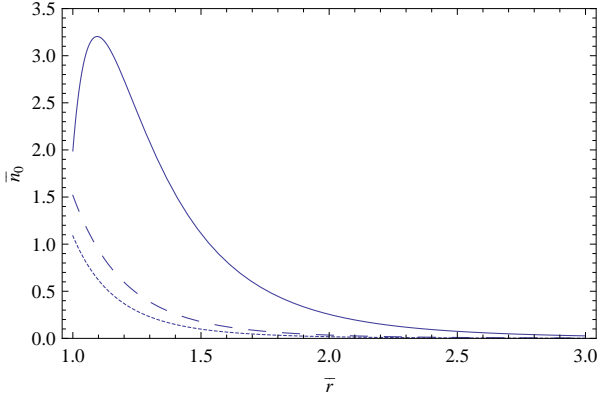} \caption{Equilibrium plasma density at the equatorial disc of the pulsar magnetosphere under gravity: solid line (---)for $\epsilon=.7$; broken line ($-$ $-$)for $\epsilon=.5$; dot line (...)for $\epsilon=.1$}
\end{figure}

The above figure (Fig.[7]) shows that the equilibrium plasma density varies significantly for increasing values of compactness parameter $\epsilon$. Very close to the surface of the compact object, for a higher value of the compactness parameter  $\epsilon=.7$, the equilibrium plasma density sharply increases  but away from the surface it falls steeply as in the Newtonian case. \\ \\

{\bf{3 Electrostatic modes in the equatorial region}}\\ \\

Against the equilibrium, now we study the low frequency (in comparison to electron cyclotron frequency
$\omega_c={eB_0}/m_ec$, where $e$ is the charge of electron, $m_e$ is the mass of electron, $c$ is the speed of
light ) electrostatic oscillations in the equatorial region ($\theta=\pi/2$). In this case, the perpendicular ( to $\widehat\theta$ ) components of electron/positron velocities in the electrostatic field $E_\perp=-\nabla_\perp\Phi$, where $\Phi$ is the electrostatic potential, are:

\begin{eqnarray}
{\bf u}_{j\perp}\approx\frac{c}{B_0(r)}\widehat{\theta}\times\nabla_\perp\Phi+\frac{cT_j}{q_jB_0(r)n_0(r)}\widehat{\theta}\times\nabla_\perp n_{j1},\label{beq26}
\end{eqnarray}
where $q_j$ is the charge of species $j=(e,p)$ and $n_{j1}(<<n_0(r))$ are small electron/positron number density perturbations.

The number density perturbations of the plasma fluids are determined by the continuity equation
\begin{eqnarray}
\frac{\partial n_{j1}}{\partial t} +\nabla_\perp\cdot\left[n_0(r)\bf{u}_{j\perp}\right]=0,\label{beq27}
\end{eqnarray}
which using the Eq. (26) can be written as

\begin{eqnarray}
\frac{\partial n_{j1}}{\partial t} &+& \frac{1}{r^2}\frac{\partial}{\partial r}\left(r\frac{cT_j}{q_jB_0(r)}\right)
\frac{\partial n_{j1}}{\partial \phi} \nonumber\\
&=& -\frac{1}{r^2}\frac{\partial}{\partial r}\left(r\frac{cn_0(r)}{B_0(r)}\right)
\frac{\partial\Phi}{\partial \phi}. \label{beq28}
\end{eqnarray}

Since, in the present study we are interested only in the toroidal modes  , we consider $n_{j1}$and $\Phi$ to be
proportional to $\exp{(-i\omega t+im\phi)}$, where $m$ is an integer $0, \; \pm 1,\; \pm 2, ....$ which represents
the toroidal mode number and $\omega$ is the frequency of the electrostatic  mode. Then, from Eq.(28), we get

\begin{eqnarray}
n_{j1}={\frac{\frac{m}{r^2}c\frac{\partial}{\partial r}(r\frac{n_0(r)}{B_0(r)})}{\omega-\frac{m}{r^2}
\frac{\partial}{\partial r}(r\frac{cT_j}{q_jB_0(r)})}}\Phi. \label{beq29}
\end{eqnarray}

From the Poisson's equation
\begin{eqnarray}
\nabla^2_\perp\Phi=4\pi e(n_{e1}-n_{p1}),\label{beq30}
\end{eqnarray}
we find the dispersion relation
\begin{eqnarray}
\omega^2 &=& \frac{m^2}{r^2}\left(\frac{1}{r}\frac{\partial}{\partial r}\left(r\frac{cT}{eB_0(r)}\right)\right)^2 \nonumber\\
&+& \frac{8\pi ec}{r^2}\frac{\partial}{\partial r}\left(r\frac{n_0(r)}{B_0(r)}\right)\frac{\partial}{\partial r}
\left(r\frac{cT}{eB_0(r)}\right),\label{beq31}
\end{eqnarray}
where, we put $T_e=T_p=T=T_0/2$.\\

Now defining

$\frac{1}{r}\frac{\partial}{\partial r}\left(r\frac{cT}{eB_0(r)}\right)=
\frac{1}{r}\frac{\partial}{\partial r}\left(r\frac{V_T^2}{\omega_c}\right)\equiv V_B$
with $V_T=\left(\frac{T}{m_e}\right)^{1/2}$ and defining $\omega_p=\left(8\pi e^2 n_0/m_e\right)^{1/2}$
for electron positron plasma, we can rewrite the dispersion relation (31) as
\begin{eqnarray}
\omega^2-K_\phi^2V_B^2+\frac{\omega_p^2}{\omega_c}K_{nb}V_B=0, \label{beq32}
\end{eqnarray}
where $K_\phi^2=\frac{m^2}{r^2}$ and $K_{nb}=-\frac{\omega_c}{\omega_p^2}\frac{1}{r}\frac{\partial}{\partial r}(r\frac{\omega_p^2}{\omega_c})$.\\ \\

{\bf Letting $\omega=\omega_r+i\Gamma$ in Eq.(32), we get}
{\bf \begin{eqnarray}
\omega_r^2-\Gamma^2-K_\Phi^2V_B^2+\frac{\omega_p^2}{\omega_c}K_{nb}V_B=0  , \label{beq33}
\end{eqnarray}}
{\bf \begin{eqnarray}
2\omega_r\Gamma=0  , \label{beq34}
\end{eqnarray}}
{\bf Eq.(34) gives two cases, namely, either}  a){\bf $\omega_r=0$ }or b){\bf $\Gamma=0.$}
{\bf For $\omega_r=0$ from Eq. (33),} we obtain the growth rate
\begin{eqnarray}
\Gamma = (\frac{\omega_p^2}{\omega_c}K_{nb}V_B -K_\phi^2V_B^2)^{1/2}  , \label{beq35}
\end{eqnarray}

which shows the electrostatic modes grow if
\begin{eqnarray}
K_{nb}>K_\Phi^2\frac{\omega_c V_B}{\omega_p^2}  . \label{beq36}
\end{eqnarray}
For a detail investigation, let us consider the o-mode i.e. $m=0$. In this case $K_\Phi=0$, then, we find
\begin{eqnarray}
\Gamma=\frac{c}{r}\sqrt{\frac{ n_0(r)T}{B_0^2(r)/8\pi}\lambda (r)}  , \label{beq37}
\end{eqnarray}
where
\begin{eqnarray}
\lambda(r)=(\frac{r}{L_B}-\frac{r}{L_n}-1)(1-\frac{r}{L_B})  , \label{beq38}
\end{eqnarray}
with
$L_B=(\frac{1}{B_0}\frac{\partial B_0}{\partial r})^{-1}$ and $L_n=(\frac{1}{n_0}\frac{\partial n_0}{\partial r})^{-1}$
representing the scale lengths of inhomogeneities of the magnetic field and the equilibrium plasma density, respectively. In fact , $\lambda (r)$ represents the combined effect of inhomogeneities.\\ \\

In Eq.(37), $\frac{n_0(r)T}{B_0^2/8\pi}$ represents the ratio of thermal pressure to the magnetic pressure, which is a positive quantity depending on $r$. Therefore, for $\Gamma >0$, $\lambda (r)$ is to be positive.  The graphical representation of $\lambda (r)$ for a particular value of $\epsilon=.7$ is shown in Fig.[8]
which shows that $\lambda (r)$ is positive for approximately $r>1.1R$ , it is negative for $r<1.1R$ and {\bf for $r \gg R$, it is saturated. The saturation is due to the fact that for $r\gg R$, $\frac{r}{L_B}\sim -3$,$\frac{r}{L_n}\sim -6$ and as a whole $\lambda (r) \sim 8$ which is a constant.}  Thus in the case of high compactness of a pulsar, the generated electrostatic oscillations in the pair plasma at the equatorial region may grow. {\bf For $\Gamma=0$, from Eq. (33), we obtain the frequency of oscillations which is given by}

\begin{figure} \vspace{6.5cm}
\includegraphics{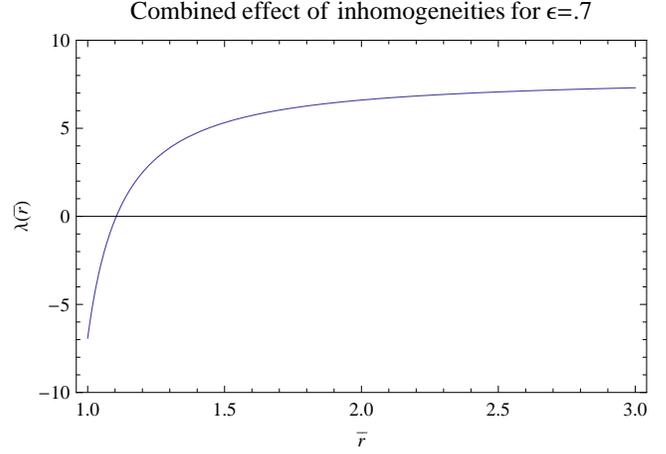} \caption{Combined effects inhomogeneities of magnetic field and plasma equilibrium density  at the equatorial disc of the pulsar magnetosphere for the compactness parameter $\epsilon = 0.7.$ }
\end{figure}
\begin{eqnarray}
\omega=\frac{c}{r}\sqrt{\frac{ n_0(r)T}{B_0^2(r)/8\pi}(1-\frac{r}{L_B}+\frac{r}{L_n})(1-\frac{r}{L_B}) }. \label{beq37}
\end{eqnarray}
{\bf which shows that $\omega\sim\frac{c}{r}(\frac{n_0(r)T}{B_0^2(r)/8\pi})^{1/2}$, where $n_0(r)T$ is the thermal pressure and $B_0^2(r)/8\pi$ is the magnetic pressure. Thus, it seems that the coupling between the thermal pressure and the magnetic pressure, which creates the  electrostatic oscillations in the pulsar magnetosphere. Although the frequency of oscillations is much lower in comparison to electron cyclotron frequency which is extremely high for a pulsar magnetosphere, in fact, a reasonably high frequency electrostatic oscillations may be created through this coupling. This coupling also may create high frequency electromagnetic oscillations, which needs further scrutiny.}

At $r=R$, we find $\omega=\omega_R\sim\frac{c}{R}\sqrt{\frac{ n_0(R)T}{B_0^2(R)/8\pi}}$. For a pulsar with super strong magnetic field, we may consider that magnetic pressure is much higher than the thermal pressure i.e. $B_0^2(R)/8\pi\gg n_0(R)T$ near the surface, then $\omega_R\ll \frac{c}{R}$. For crab pulsar PSR 0531 +21, $B_0(R)=10^{13}G $ and $R=10^6 cm$ (Nanobashvili, 2004). So we find $\omega_R\ll 3\times 10^4 s^{-1}$ and $\omega_c\sim 10^{20} s^{-1}$, which validate our approximation $\omega \ll \omega_c$ {\bf and accordingly we ignored the effects of currents and conduction. } \\
{\bf So far, we consider the nonaligned magnetic field. It would be interesting to investigate the growing modes in oblique direction of the magnetic field. To find clues in this case,we have investigated  the simple case of aligned magnetic field (Michel and Li, 1999), which shows the mode grows from a particular range of angle inclination $\alpha > sin^{-1}\sqrt{\frac{16}{5}\frac{n_pT}{B_p^2/8\pi}}$.}\\

{\bf{4 Discussion}}\\
Here,we study the low frequency (in comparison to the cyclotron frequency)electrostatic modes in the isothermal electron-positron plasma of the pulsar magnetosphere. The compactness of the pulsar is considered in this investigation. The equilibrium magnetic field and the equilibrium plasma density are investigated  in detail. To visualize the picture of the magnetic field around pulsar, the magnetic lines of force are drawn for both the Newtonian case (no gravity) and for the compactness case (with gravity). The corresponding magnetic field distribution in the meridional plane are shown in  density plots. The equilibrium is due to the balance of thermal pressure with the magnetic pressure. In the Newtonian case, it is found that the equilibrium plasma density is isotropic and is steeply decreasing function of radial distance. But for the compact pulsar with gravity, the equilibrium plasma density varies significantly for increasing values of compactness parameter. Further, against the equilibrium, we study plasma perturbation in the linear approximation.  The number density perturbations of the plasma fluids are determined by the continuity equation, which coupled with the Poisson's equation, yields the dispersion relation for the electrostatic modes.  From the analysis of the dispersion relation, we find  growing modes, which are related with the inhomogeneity scales of equilibrium magnetic field and the equilibrium plasma density. The thermal pressure coupled with the magnetic pressure creates the low frequency electrostatic modes  which propagate along the toroidal direction of the pulsar.\\

{\bf{5 Conclusion}}\\
Growing elecrostatic modes may be generated in the equatorial region of a compact pulsar magnetosphere. The modes are created due to the coupling of the thermal pressure with the magnetic pressure, which may grow for the particular cases of inhomogeneities of the equilibrium magnetic field and the equilibrium plasma density. This may have some relevance to the recent discoveries of low frequency quasi-periodic oscillations (QPOs) of magneter with an extremely strong magnetic field (Lee, 2008).\\

{\bf{ Acknowledgement}}\\

This work has been supported financially by the Ministry of Education of  Government of Bangladesh under Grants for Advanced Research in Science:MOE.ARS.PS.2011. No.-86 .\\ \\

{\bf{ References}}\\ \\
\noindent
Arons, J., Scharlemann,E. T.:Pair formation above  pulsar polar caps - Structure of the low altitude\indent acceleration zone, ApJ, 231, 854 (1979)\\ \\
Banarjee, D., Bhatt, J. R., Das, A. C., Prasanna,\indent A. R.: Structure of a fluid disk around a magnetized compact object in the presence of a self consistent toroidal magnetic field, ApJ, 449, 789 (1995)\\ \\
Buzzi,V.,Hines,K.C.:Relativistic plasmas near a\hspace{1.5cm}Schwarzschild black hole: A solution of the two fluid ODE's in Schwarzschild coordinates,Phys. Rev.D,51,6692 (1995)\\ \\
Ginzburg, V. L., Ozernoy, L. M.: Magnetic Models of Pulsars and Rotating Neutron Stars, Zh. Eksp. Teor. fiz., 47, 1030 (1964)\\ \\
Goldreich, P., Julian, W. H.:Pulsar electrodynamics, ApJ, 157, 869 (1969)\\ \\
Krall, N. A., Trivelpiece, A. W.: Principles of Plasma Physics, Mc Graw-Hill Company, 1973\\ \\
Lee, U.: Axisymmetric toroidal modes of magnetized neutron star, MNTRS, 385, 2069 (2008)\\ \\
Lominadze, J. G., Machabeli, G. Z., Usov, V. V.: Theory of NP032 pulsar radiation and the nature of the activity of the Crab Nebula, Astrophys. Space Sci., 90, 19 (1983).\\ \\
{\bf Michel, F. C., Li, H.: Electrodynamics of neutron stars,Physics Reports, 318,254 (1999)}\\ \\
Mofiz U. A.: Isolated solitons in an ultrarelativistic electron-positron plasma of a pulsar magnetosphere.- Phys. Rev. A 40, 2203 (1989)\\ \\
Mofiz, U. A.: Ultrarelativistic envelope solitons in a magnetized electron-positron plasma
,Phys. Rev. A, 42, 960 (1990)\\ \\
Mofiz,U.A.: Linear modes in the rotating neutron star polar-cap electron-positron plasma, Phys. Rev. E, {\hspace{.5cm}} 55,5894 (1997)\\ \\
Mofiz, U. A., De Angelis, U., Forlani, A.: Solitons in weakly nonlinear  electron-positron plasmas and pulsar microstructure, Phys. Rev. A 31, 951(1985)\\ \\
Mofiz, U. A., Podder, J.: Solitons in strongly magnetized electron-positron plasma and pulsar microstructure, Phys. Rev. A 36, 1811 (1987)\\ \\
Mofiz, U. A., Ahmedov, B. J.: Plasma modes along the open field lines of a neutron star, ApJ, 542, 484 (2000)\\ \\
Muslimov,A., Harding, A.K.:Toward the quasi-steady state electrodynamics of a neutron star, ApJ, 485, 735 (1997)\\ \\
Nanobashvilli, I. S.: Toroidal magnetic field generation in the magnetosphere of crab pulsar. Astrphys. and Space sci. 294, 125 (2004)\\ \\
Prasanna, A. R., Tripathy, S. C., Das, A. C.: Equlibrium structure for a plasma magnetosphere around compact objects, J. Astrophys. Astr. 10, 21 (1989)\\ \\
Ruderman,M.,Sutherland,P.G.:Theory of pulsars - Polar caps, sparks, and coherent microwave radiation \hspace{.5cm}ApJ,196,51 (1975)\\ \\
Sakai,J., Kawata,T. J.: Nonlinear Alfve'n wave in ultra-relativistic electron-positron plasma, J. Phys. Soc. Japan, 49, 753 (1980)\\ \\
Sana, A., Mendez, M., Altamirano, D., Homan, J.,Casella, P.,Belloni, T., Lin, D., klis,, M., Wijhands, R.:The kilohertz quasi-periodic oscillations during the z and attoll phases of the unique transient XTE J1701-62, MNTRS, 1365 (2010)\\ \\
Shapiro, S. L., Teukolsky, S. A.: Black Holes, White Dwarfs and Neutron Stars, Wiley-VCH, 2004, p. 277\\ \\
Shukla, N., Shukla, P. K.: A new purely growing instability in a strongly magnetized nonuniform pair plasma, Phys. Lett. A, 367, 120 (2007)\\ \\
Sturrock,P.A.,Baker, K. B. :Positron production by pulsar, ApJ, 234,612 (1979)\\ \\
Usov, V. V.; Melrose, D. B. :Pulsars with strong magnetic fields - polar gaps bound pair creation and nonthermal luminosities,Australian Journal of Physics,48, 571 (1995)\\

\end{document}